%% file: main.tex
\documentclass[twocolumn]{ceurart}

\usepackage{listings}
\usepackage[normalem]{ulem}
\usepackage{xspace}

\newcommand{\sqlite}{\textrm{SQLite}\xspace}
\newcommand{\postgres}{\textrm{PostgreSQL}\xspace}
\newcommand{\bigquery}{\textrm{BigQuery}\xspace}
\newcommand{\duckdb}{\textrm{DuckDB}\xspace}

\input{logica_style}

\usepackage[framemethod=tikz]{mdframed}

\begin{document}


\copyrightclause{for this paper by its authors. Use permitted under Creative Commons License Attribution 4.0 International (CC BY 4.0).}

\conference{Published in the Proceedings of the Workshops of the EDBT/ICDT 2025 Joint Conference (March 25-28, 2025), Barcelona, Spain}

\title{Logica-TGD: Transforming Graph Databases Logically}

\author[1]{Evgeny Skvortsov}[%
email=evgenys@google.com,
]
\address[1]{Google LLC, WA, USA}

\author[2]{Yilin Xia}[%
email=yilinx2@illinois.edu,
]
\author[2]{Bertram Lud\"{a}scher}[%
email=ludaesch@illinois.edu,
]
\address[2]{University of Illinois Urbana-Champaign, School of Information Sciences, IL, USA}

\author[3]{Shawn Bowers}[%
email=bowers@gonzaga.edu,
]
\address[3]{Gonzaga University, Department of Computer Science, Spokane, WA, USA}

\begin{abstract}
Graph transformations are a powerful computational model for manipulating complex networks, but handling temporal aspects and scalability remain significant challenges. We present a novel approach to implementing these transformations using Logica, an open-source logic programming language and system that operates on parallel databases like DuckDB and BigQuery. Leveraging the parallelism of these engines, our method enhances performance and accessibility, while also offering a practical way to handle time-varying graphs. We illustrate Logica's graph querying and transformation capabilities with several examples, including the computation of the well-founded solution to the classic ``Win-Move'' game, a declarative program for pathfinding in a dynamic graph, and the application of Logica to the collection of all current facts of Wikidata for 
taxonomic relations analysis. We argue that clear declarative
syntax, built-in visualization and powerful supported engines make
Logica a convenient tool for graph transformations.

\end{abstract}

\begin{keywords}
  Logic rules \sep graph queries \sep
  graph transformations 
\end{keywords}

\maketitle

\section{Introduction}

Graph transformations are a powerful and versatile method for modeling and manipulating complex systems across diverse fields, ranging from software engineering \cite{Parra14,Krauter23} and social network analysis \cite{Fernandez16} to biology and chemistry \cite{RosselloBio2005,Nagl76,Rossello2Chem005}. These transformations typically operate by applying rewrite rules to a graph, altering its structure and properties. While these transformations are known for their expressiveness and flexibility, their implementation can often be complex, especially when dealing with time-varying graphs or requiring scalable solutions. Existing graph database systems often provide limited support for such evolution mechanisms, creating a gap between theoretical models and practical implementations.

\sloppypar Logic programming, on the other hand, provides a declarative approach to problem solving by expressing rules and relationships rather than explicitly stating control-flow. Logic-based systems (e.g., Prolog and Answer Set Programming  \cite{Gelfond08}) are well-established in various areas, including (symbolic) AI and knowledge representation and reasoning. More recently, \href{https://logica.dev/}{Logica} (\uline{\emph{Logic}} \texttt{+} \uline{\emph{A}}ggregation) \cite{Skvortsov_EDBT_2024,Skvortsov_Datalog2_2024}, an open-source logic programming language and system, has emerged, which employs parallel data processing environments such as \href{https://duckdb.org/}{DuckDB} and \href{https://cloud.google.com/bigquery}{BigQuery}. Logica combines the declarative power of logic programming with the scalability and efficiency of modern databases, offering a promising new path for tackling graph transformation challenges.

This paper explores the application of Logica to the domain of graph transformations, bridging the gap between these two paradigms. Our approach leverages Logica's ability to process large datasets in a parallel and efficient manner, providing a novel means of implementing and executing graph transformations at scale using logical rules. We demonstrate that with a graph transformation mindset, logic programming can provide a natural, powerful, and intuitive approach for defining transformations, 
opening new opportunities within both communities. Crucially, Logica also enables a practical approach to addressing issues that are difficult in classical graph transformation approaches, such as temporal transformations and seamless scalability.

To illustrate the practical nature of our approach, we introduce several example Logica programs implementing different types of graph transformations. Specifically, we show how Logica can be used to define and perform basic graph transformations including message passing, removing transitively implied graph edges, and solving (i.e., labeling or coloring) win-move graphs.  
We also introduce a novel approach to time-aware pathfinding (e.g., \cite{kempe_2002,Rossello2005,Casteigts_2021}), directly addressing the need for principled and tractable mechanisms to handle evolving graph data, a notoriously difficult problem in classical graph transformation approaches. These examples serve both to show the expressiveness of our approach, and to illustrate its potential benefits.

The rest of this paper is structured as follows: Secion 2 provides an overview of Logica. Section 3 demonstrates how to express graph transformations in Logica. We conclude in Section 4 with a discussion of the limitations and potential avenues for future research.

\section{Logica Overview}

Logica is a freely available, open-source variant of Datalog combining the declarative features of
expressive rule-based languages with aggregation. Logica is a descendant of Yedalog
\cite{chin2015yedalog} and inherits several of its features including support for aggregators,
functional predicates, user-defined functions, and complex data types. Logica also supports rules
involving recursion and negation.  A key feature of the implementation is that it converts programs
into SQL. This conversion can be configured by the user and rules can be compiled into: (a)
self-contained SQL scripts with fixed recursion depth; or (b) Python-driven pipelines that chain
together SQL queries when deep recursion is needed.

\begin{figure*}[!t]
  \centering\includegraphics[width=0.75\textwidth]{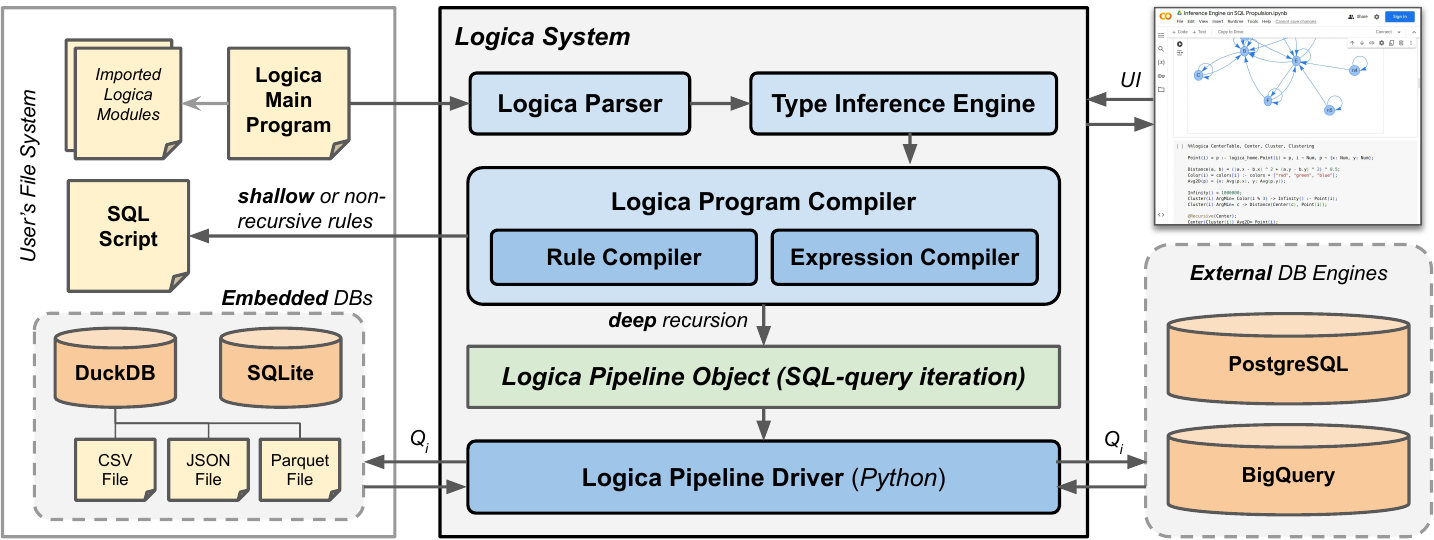}
  \caption{{\footnotesize\textbf{Logica System Architecture (from \cite{Skvortsov_Datalog2_2024})}: Logica supports a module system for importing libraries (top left); users  write programs locally (top left) or from within Jupyter notebooks (top right); programs are parsed, type checked (top middle), and compiled to a self-contained SQL script (no/limited recursion) or to an iterative plan of pipelined SQL-queries passed to a pipeline driver (bottom middle). Logica executes and monitors queries as they run on underlying SQL engines, either locally (bottom left) or externally (bottom right).}}
  \label{fig:system}
\end{figure*}

An overview of the Logica system is shown in Figure~\ref{fig:system}. Developers can work with
Logica from the command line or via a Jupyter notebook. Programs can import functions and other
rules via a module system. Logica parses and analyzes program files and produces a collection of SQL
queries in the dialect of the target database engine (currently \sqlite, \duckdb, \postgres,
or \bigquery). To help manage differences among platforms, Logica employs a type inference engine to
create correct SQL for each underlying system. If the set of rules to be evaluated is
non-recursive (or if the user has indicated via a directive that a fixed-depth, non-iterative
recursion is sufficient), a self-contained SQL script is generated that can be directly
executed. For programs requiring deep recursion, Logica generates a pipeline script that iteratively
executes the generated SQL queries stage-by-stage until a fixpoint or a user-defined termination
condition is reached.

For long-running queries, users can monitor rule execution in the Logica UI: predicate results are
rendered as they are being evaluated, so the user knows which (iterated) relations are still
running. This information can also be saved and used for logging and profiling program execution.

\section{Transforming Graphs with Logica}

In Logica, as in Prolog and Answer-Set Programming, graphs can be naturally represented: Nodes are denoted using unique identifiers and edges are represented as facts about the nodes. Graphs can be represented as predicates with two or more \emph{positional} arguments. For example, a simple directed graph can be represented as a binary relation \texttt{E(}\emph{source}\texttt{,}\emph{target}\texttt{)}, where \emph{source} and \emph{target} are nodes connected by a directed edge. Additional graph data, e.g., node or edge attributes, can be expressed using
separate predicates or extra {positional} (in Logica also \emph{named}) arguments. For instance, a graph whose edges are assigned colors can be represented using a ternary predicate and facts of
the form \texttt{E(}\emph{source}\texttt{,}\emph{target}\texttt{,}\emph{color}\texttt{)}.

Graph transformations can be implemented by defining new predicates that depend on the original
graph: e.g.,  given a graph with edges
\texttt{E}, a new graph \texttt{E2} extending \texttt{E} by adding new edges between nodes that are
separated by two ``hops''  can be derived as follows.\footnote{In Logica, variables are lowercase, predicates are uppercase, ``$\sim$'' is used for negation, and semicolons denote the end of rules.} 
\begin{mdframed}[backgroundcolor=gray!8!white,hidealllines=true, skipabove=0.2cm, innertopmargin=-0.0cm,innerbottommargin=-0.0cm]
\begin{lstlisting}[style=logica]
  E2(x, z) :- E(x, y), E(y, z);
  E2(x, y) :- E(x, y);
\end{lstlisting}
\end{mdframed}
This simple example demonstrates an important aspect of defining graph
transformations with logic rules. In native graph transformation languages
users define how the graph changes, and edges not involved in the change
remain. However, when defining graph transformations logically, we must include rules to
\emph{preserve} edges not involved in the transformation (line 2).

\subsection{Message Passing}

Our first example involves the simple passing of a message along the
directed edges of a graph, as described in \cite{konig_graph_transformation_2020}.

\begin{mdframed}[backgroundcolor=gray!8!white,hidealllines=true, skipabove=0.2cm, innertopmargin=-0.0cm,innerbottommargin=-0.0cm]
\begin{lstlisting}[style=logica]
M0(0);                  # start node
# Rule 1: Message initialization.
M(x) :- M=nil, M0(x);   
# Rule 2: Message passing.
M(y) :- M(x), E(x, y);  
# Rule 3: Message retention.
M(x) :- M(x), ~E(x, y); 
\end{lstlisting}
\end{mdframed}

\noindent
Fact  \texttt{M0(0)} states that initially the message is at node \texttt{0}. Rule 1 initializes the message relation \texttt{M}. It states that node \texttt{x} has a message \texttt{M} if it was specified as the initial node in \texttt{M0(x)}. Condition \texttt{M=nil} makes the
rule fire (trigger) only at the start of the recursive iteration. The message propagates from node \texttt{x} to its neighbors \texttt{y} through the Rule 2. Finally, Rule 3 requires the message of a node \texttt{x} to be retained if \texttt{x} has no outgoing edges. 

\subsection{Computing Distances in a Graph}

Logica has native support for aggregation. Predicate level aggregation
is computed over the specified values of the fields of the defined predicate. 
The following Logica program computes the minimum distance \texttt{D(x)} of a node \texttt{x} from a given \texttt{Start} node.

\begin{mdframed}[backgroundcolor=gray!8!white,hidealllines=true, skipabove=0.2cm, innertopmargin=-0.0cm,innerbottommargin=-0.0cm]
\begin{lstlisting}[style=logica]
# Rule 1: Distance from the Start node is 0.
D(Start()) Min= 0;
# Rule 2: Triangle inequality.
D(y) Min= D(x) + 1 :- E(x,y);
\end{lstlisting}
\end{mdframed}

\sloppypar \noindent 
\texttt{Start()} is used as a \emph{functional predicate}: All Logica relations have an additional ``special attribute'' (named \texttt{logica\_value}) to store and access a relation's functional value, i.e., its value when used syntactically as a function. Here, \texttt{Start()} simply returns (a predefined) starting value.  Rule 1 assigns the distance \texttt{0} to the starting node, where \texttt{D(x)} is also a functional predicate that returns the (minimum) distance of a node \texttt{x}. Rule 2 computes the distance to a node \texttt{y} as {\em at most} one more than the distance to its predecessor \texttt{x}. The \texttt{Min=} aggregation operator specifies that the minimum distance will be taken among all possible paths.

\subsection{Solving Win-Move Games}
Win-Move is a two-player combinatorial game played on a finite directed graph (representing a board) where nodes represent positions and edges represent moves \cite{smith_graphs_1966,FKL97}. After choosing a starting position (node), the two players take turns moving a game pebble along the edges of the graph. A player who cannot move loses the game. We propose to compute the solution to Win-Move games in a new way, inspired by graph transformations. A concise formulation of the game is given by the logic rule:
$$\verb|Win(x) :- Move(x,y), ~Win(y).|$$

A solution to the game can be computed using the 3-valued  well-founded semantics \cite{van_gelder_well-founded_1991}.  The moves of the game are given by the predicate \verb|Move|. A game starting at node \verb|x| is objectively won if there exists a (\emph{winning}) move to some \verb|y| that is lost. Then, no matter how an opponent (Player II) moves from \verb|y| (assuming there is an outgoing move to begin with), Player I can always force a win after the opponent has moved. The well-founded model assigns \verb|Win(x)| the value \emph{true}, \emph{false}, and \emph{undefined} if and only if \verb|x| is objectively \emph{won}, \emph{lost}, and \emph{drawn}, respectively.
In contrast, the 2-valued \emph{stable model} semantics does not compute the desired game solution: There can be 0, 1, or more stable models, none of which may agree with the correct, well-founded solution.
Similarly, in Logica, the above rule cannot be used as-is.

Looking at the problem through a graph-transformation lens, however, we
can define a new predicate \verb|W(x,y)|, indicating that the move from \verb|x| to \verb|y| is a winning move. 
Computation of  $\verb|W|$, from a graph transformation perspective,
amounts to selecting a collection of winning moves. In Logica this can be specified with a single rule:

\begin{mdframed}[backgroundcolor=gray!8!white,hidealllines=true, skipabove=0.2cm, innertopmargin=-0.0cm,innerbottommargin=-0.0cm]
\begin{lstlisting}[style=logica]
W(x,y) :- Move(x,y), (Move(y,z1) => W(z1,z2));
\end{lstlisting}
\end{mdframed}
The rule says: A move from \verb|x| to \verb|y| is \emph{winning} iff for \emph{every} opponent move from \verb|y| to some \verb|z1| there \emph{exists} a move from \verb|z1| to \verb|z2| for Player I that is again winning! The logical implication $\verb|Move(y,z1) => W(z1,z2)|$ is a shorthand for the nested negation $\verb|~(Move(y,z1), ~W(z1,z2))|.$

The position values can now be derived from the winning moves \verb|W|. 
The source and target nodes \verb|x| and \verb|y| of a winning move are won and lost positions, respectively. Positions that are neither won nor lost are drawn:
\begin{mdframed}[backgroundcolor=gray!8!white,hidealllines=true, skipabove=0.2cm, innertopmargin=-0.0cm,innerbottommargin=-0.0cm]
\begin{lstlisting}[style=logica]
Won(x), Lost(y)  :- W(x,y);
Drawn(x) :- Position(x), ~Won(x), ~Lost(x);
\end{lstlisting}
\end{mdframed}
Here, the unary predicate \verb|position| is  defined as the union of the domain and range of the \verb|Move| relation: 
\begin{mdframed}[backgroundcolor=gray!8!white,hidealllines=true, skipabove=0.2cm, innertopmargin=-0.0cm,innerbottommargin=-0.0cm]
\begin{lstlisting}[style=logica]
Position(x) :- x ~in~ [a,b], Move(a,b);
\end{lstlisting}
\end{mdframed}

\subsection{Finding Paths in Evolving Graphs} 

Graphs that change over time \cite{kempe_2002,Rossello2005,Casteigts_2021} can be naturally modeled and computed over in Logica, where 
time can represented using extra positional arguments, named arguments, or
as functional values. As a simple example, assume a graph that changes 
over time is represented by the relation \texttt{E(x,y,t0,t1)}
stating that there exists an edge from node \texttt{x} to
\texttt{y} added to the graph between time moments \texttt{t0} and \texttt{t1}.
For simplicity we assume that time starts at moment 0, edges are transitioned instantly, and the start node is given as \texttt{Start()}.
The following Logica program computes the earliest possible arrival time for
each node.

\begin{mdframed}[backgroundcolor=gray!8!white,hidealllines=true, skipabove=0.2cm, innertopmargin=-0.0cm,innerbottommargin=-0.0cm]
\begin{lstlisting}[style=logica]
# Rule 1: Starting condition.
Arrival(Start()) Min= 0;
# Rule 2: Traversal of an edge when edge exists.
Arrival(y) Min= Greatest(Arrival(x),t0) :-
  E(x,y,t0,t1), Arrival(x) <= t1;
\end{lstlisting}
\end{mdframed}

\begin{figure*}[t!]
  \centering
  \includegraphics[width=0.85\linewidth]{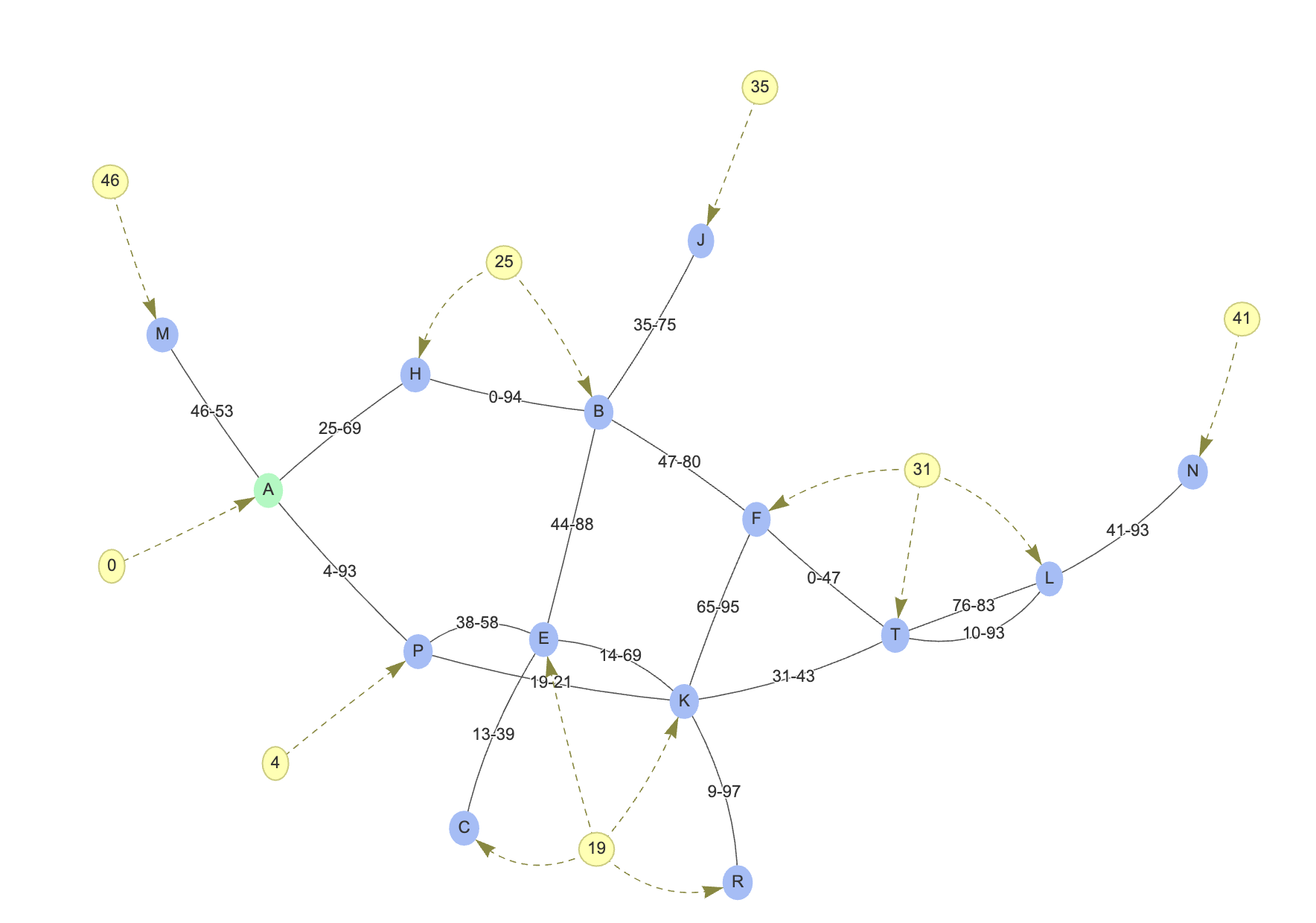}
  \caption{Visualizing the solution of pathfinding in a dynamic graph. The time of existence is shown as labels
  on edges; the earliest possible time of arrival is shown via yellow nodes.  Section~\ref{subsection:effortless} shows how such visualizations can be created in Logica.
  }
  \label{fig:arrival-time}
\end{figure*}

Rule 1 sets the arrival time of the start node to 0 (the assumed initial time). Rule 2 considers edge transitions: if we arrive at 
the source \texttt{x} of an edge before the edge expires (given by \texttt{t1}) then the
arrival time of the edge's target \texttt{y} is set to the larger (latest) of \texttt{x}'s arrival time and the time \texttt{t0} that the edge was added to the graph. Once the arrival times are computed, additional rules can be easily added to select specific (time aware) paths of the graph.  Figure~\ref{fig:arrival-time} gives an example of path finding over an evolving graph. An initial graph (with nodes \textsf A, \textsf B, etc., and colored blue) is given with edges labeled by their start and end times. The start node \textsf A is shown in green. The computed arrival times are displayed as additional nodes (in yellow). The graph in Figure~\ref{fig:arrival-time} was created within Logica as further described below in Section~\ref{subsection:effortless}.  





\subsection{Transitive Reduction of DAGs}
\label{subsec:TR}

The {\em transitive reduction} \cite{aho_garey_ullman_1972} of a directed graph is a graph with the fewest possible edges that has the same reachability as the original graph. For directed acyclic graphs (DAGs), the transitive reduction
coincides with finding the (unique) {\em minimum equivalent subgraph}, i.e., a proper subgraph with the fewest possible edges needed to maintain the same reachability relation as the original graph. While finding a minimum equivalent subgraph for (arbitrary) graphs with cycles is NP-complete, for DAGs, the problem can be solved in PTIME. The following Logica program computes the transitive reduction \texttt{TR(x,y)} of a DAG given by the edge relation \texttt{E(x,y)}. The program first computes the transitive closure and then uses the transitive closure to remove redundant edges.

\begin{mdframed}[backgroundcolor=gray!8!white,hidealllines=true, skipabove=0.2cm, innertopmargin=-0.0cm,innerbottommargin=-0.0cm] \begin{lstlisting}[style=logica]
# Rule 1: Transitive closure base case.
TC(x,y) distinct :- E(x,y);
# Rule 2: Transitive closure inductive step.
TC(x,y) distinct :- TC(x,z), TC(z,y);
# Rule 3: Transitive reduction.
TR(x,y) :- E(x,y), ~(E(x,z), TC(z,y));
\end{lstlisting} \end{mdframed}

Rules 1 and 2 compute the transitive closure. Rule 3 identifies edges \texttt{E(x,y)} in the original graph that cannot be bypassed by going to some other node and taking a transitive path from there. These edges are the essential edges needed to maintain reachability and make up the edges of the resulting transitive reduction graph \texttt{TR}.



\subsection{Rendering Graphs with Logica}
\label{subsection:effortless}

One of the benefits of Logica is the convenience of rendering graphs directly from predicate definitions.  With just a few lines of Logica code and a minimal Python wrapper, it is possible to generate visually appealing and informative graph representations that highlight
case-specific attributes. For example, consider the transitive reduction computed in the previous section.  Instead of exporting the \verb|TR| and \verb|E| predicates and configuring their rendering in a separate graphing library, Logica makes it possible to define a predicate that {\em directly} specifies the visual attributes of the graph. The following relations define graph visualization properties in Logica highlighting the edges in both the original graph and its transitive reduction:

\begin{mdframed}[backgroundcolor=gray!8!white,hidealllines=true, skipabove=0.2cm, innertopmargin=-0.0cm,innerbottommargin=-0.0cm] \begin{lstlisting}[style=logica]
R(x, y,
  arrows: "to",
  color? Max= "rgba(40, 40, 40, 0.5)",
  dashes? Min= true, 
  width? Max= 2,
  physics? Max= false,
  smooth? Max= false) distinct :- E(x, y);
R(x, y,
  arrows: "to",
  color? Max= "rgba(90, 30, 30, 1.0)",
  dashes? Min= false, 
  width? Max= 4,
  physics? Max= true,
  smooth? Max= true) distinct :- TR(x, y);
\end{lstlisting} \end{mdframed}

The \texttt{R(x,y,$\dots$)} relation defines the edges of the graph and their visual properties.  The first rule draws the original edges from \verb|E(x, y)| in a light gray, dashed style with a thin line. The second rule draws the transitive reduction edges from \verb|TR(x, y)| in bold red, solid lines. The \verb|distinct| keyword triggers aggregation, so that
each edge occurs only once and values of attributes such as \verb|color|,
\verb|dashes|, etc., are chosen based on whether this edge
belongs to the transitive reduction. 
The visual attributes, e.g., \verb|arrows|, \verb|color|, \verb|dashes|, \verb|width|, \verb|physics|, and \verb|smooth|, are directly embedded within the Logica rules.  The question mark in ``\verb|color? Max=|'' indicates that the \verb|color| can have different values from different rules.  The \verb|Max=| indicates that if there are multiple rules that apply to the same edge, the maximum value will be used. The actual rendering is then handled with a Python
function call:

\begin{mdframed}[backgroundcolor=gray!8!white,hidealllines=true, skipabove=0.2cm, innertopmargin=-0.0cm,innerbottommargin=-0.0cm] \begin{lstlisting}[style=logica]
# from logica.common import graph
graph.SimpleGraph(
  R, extra_edges_columns=[
    "arrows", "physics", "dashes", "smooth"], 
  edge_color_column="color",
  edge_width_column="width")
\end{lstlisting} \end{mdframed}

\begin{figure*}[t!]
  \centering
  \includegraphics[width=0.72\linewidth]{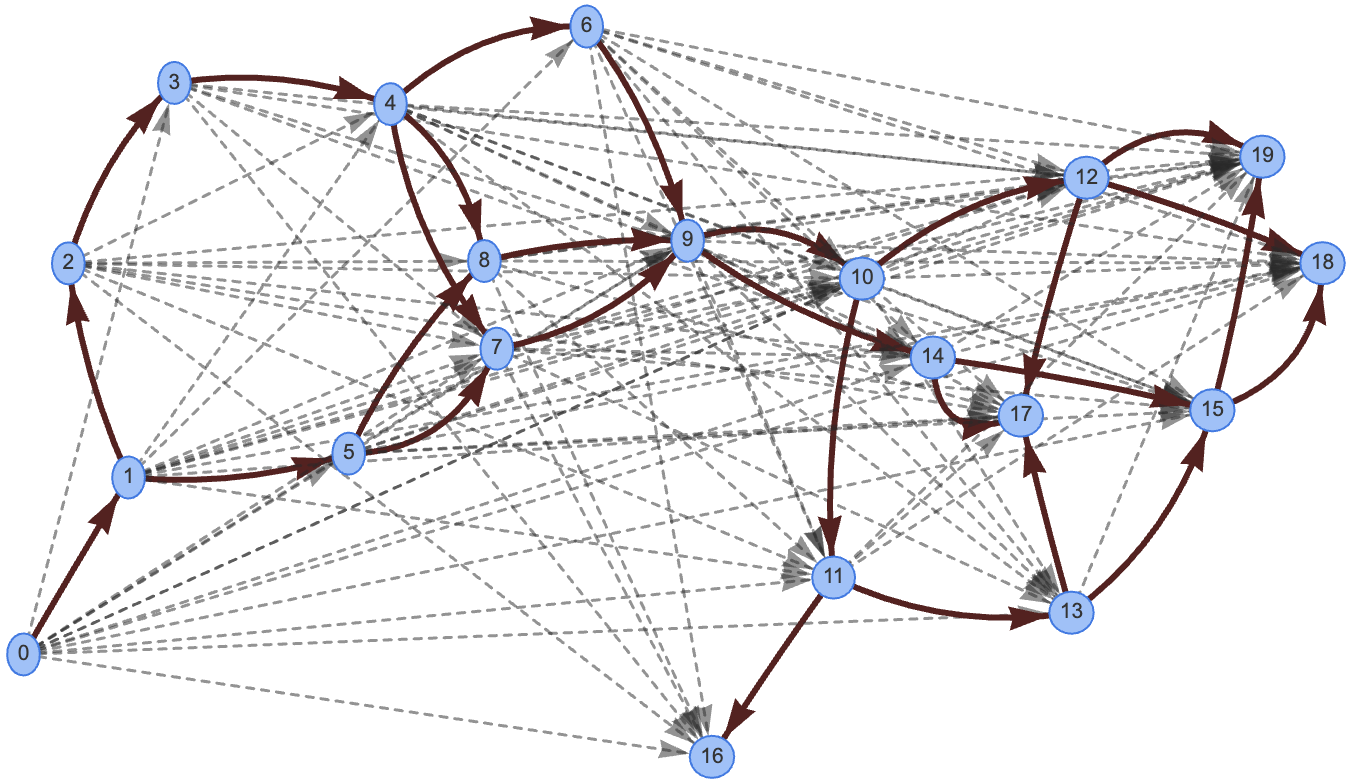}
  \caption{Visualizing a transitive reduction with Logica: Original edges are displayed with dashed lines, while edges present in the transitive reduction are highlighted with solid lines. This visualization, generated directly from Logica predicates, shows how the transitive reduction simplifies the graph while preserving reachability.}
  \label{fig:transitive-reduction}
\end{figure*}

This code leverages Logica's \verb|graph| module.  The \verb|SimpleGraph| function takes the \verb|R| predicate, specifies which columns in \verb|R| represent edge attributes (using \verb|extra_edges_columns|, \verb|edge_color_column|, and \verb|edge_width_column|), and provides overall graph options. The resulting visualization is shown in Figure~\ref{fig:transitive-reduction} (where the transitive reduction graph is overlaid over the original graph). The tight integration between logical definitions and graph rendering makes Logica a powerful tool for analyzing and understanding complex relationships in graph data. The ability to define graph properties directly in Logica, rather than relying on external tools, can significantly streamline the data exploration and visualization process. 

\subsection{Graph Condensation} 

Graph condensation \cite{tarjan_1972} is a technique for simplifying a graph by collapsing strongly connected components (SCCs) into single nodes. This can be useful for visualizing the overall structure of a graph and for performing analyses that are less sensitive to the details within SCCs. In the condensed graph, each node represents an SCC, and an edge exists between two nodes if there is an edge in the original graph between nodes in the corresponding SCCs.

The following Logica program computes the condensation of a graph given by edges defined by the \verb|E(x,y)| relation, and where all nodes are defined by a \verb|Node(x)| predicate. We assume that transitive
closure is already computed as in Subsection~\ref{subsec:TR}.


\begin{mdframed}[backgroundcolor=gray!8!white,hidealllines=true, skipabove=0.2cm, innertopmargin=-0.0cm,innerbottommargin=-0.0cm] 
\begin{lstlisting}[style=logica]
# Minimal node ID of the component 
# .. is used as the component ID.
CC(x) Min= x :- Node(x);
CC(x) Min= y :- TC(x,y), TC(y,x);
# Compute condensation graph edges.
ECC(CC(x),CC(y)) distinct :- 
  E(x,y), CC(x)!=CC(y);
\end{lstlisting} 
\end{mdframed}



Once the condensation is computed, it can be easily rendered using Logica's graph visualization capabilities. To enhance readability, we use the following naming conventions: nodes in the original graph are named directly with their IDs (e.g., ``1'', ``2'', ``3''), while the condensed components are prefixed with ``c-'' (e.g., ``c-1'', ``c-2''). The predicate definitions for rendering both the original and condensed graphs simultaneously are as follows:

\begin{mdframed}[backgroundcolor=gray!8!white,hidealllines=true, skipabove=0.2cm, innertopmargin=-0.0cm,innerbottommargin=-0.0cm] \begin{lstlisting}[style=logica]
NodeName(x) = ToString(ToInt64(x));
CompName(x) = "c-" ++ ToString(ToInt64(x));

# Edges of the original graph.
Render(NodeName(a), NodeName(b),
       physics: true, arrows: "to",
       dashes: false, smooth: true,
       color: "#33e") :- E(a, b);
# Edges of the condensation.
Render(CompName(x), CompName(y),
       physics: true, arrows: "to",
       dashes: false, smooth: true,
       color: "#33e") :- ECC(x, y);
# Mapping from the graph to
# the condensation.
Render(NodeName(ToInt64(a)), CompName(CC(a)),
       physics: false, arrows: "to",
       dashes: true, smooth: false,
       color: "#888");
\end{lstlisting} \end{mdframed}

\begin{figure*}[t!]
  \centering
  \includegraphics[width=0.55\linewidth]{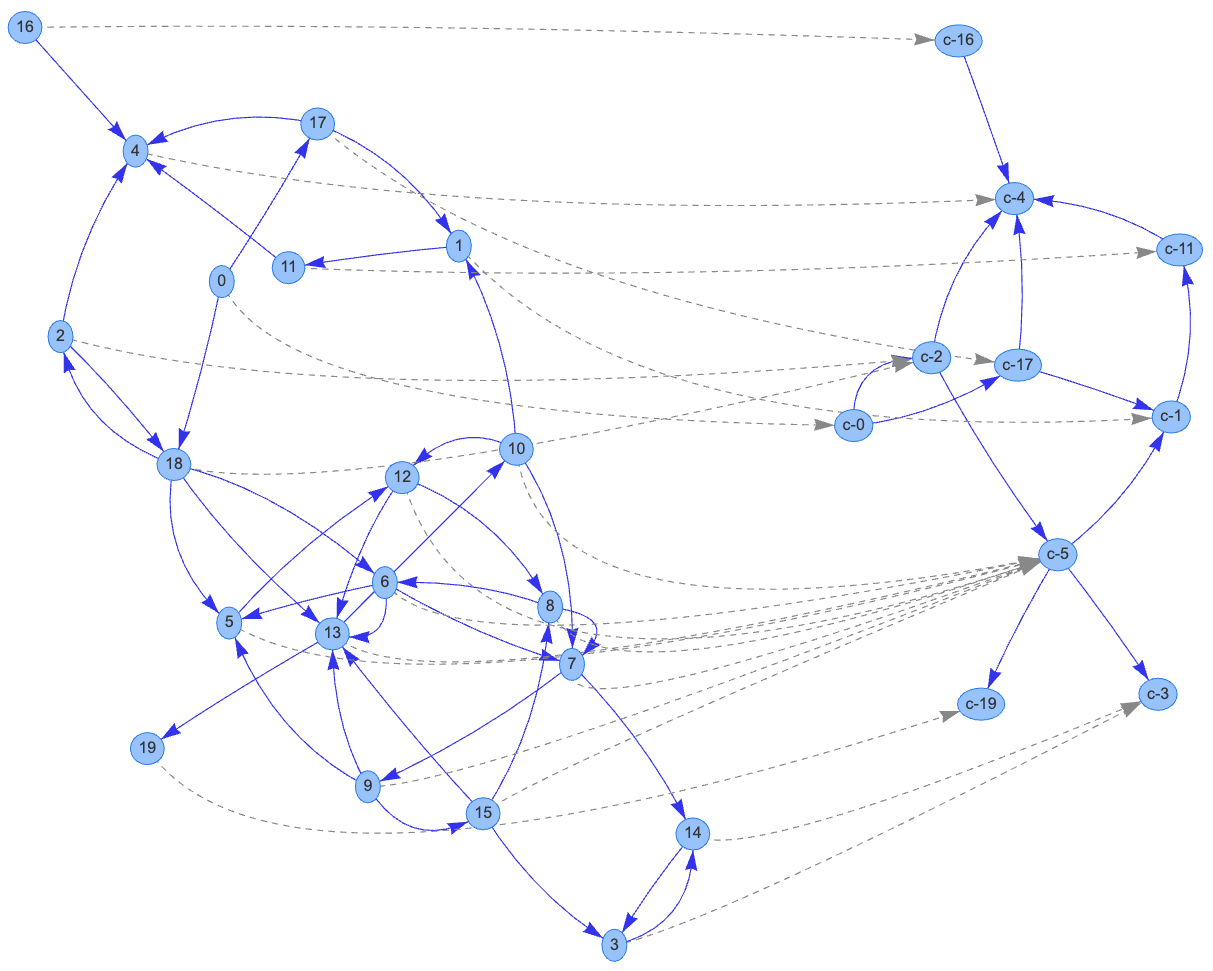}
  \caption{Graph Condensation visualized with Logica. This figure displays both the original graph and its condensed representation. Solid blue lines represent edges within the original graph and between condensed components. Dashed gray lines connect each node to its corresponding condensed component, illustrating how strongly connected components are collapsed into single nodes. This condensation simplifies the graph while preserving high-level connectivity information.}
  \label{fig:graph-condensation}
\end{figure*}

\noindent The \verb|NodeName| and \verb|CompName| functions convert node and component IDs to strings with appropriate prefixes. The \verb|Render| predicate then defines the appearance of the graph:
\begin{itemize}
\item Edges in the original graph (defined by \verb|E|) are rendered as solid blue lines.
 \item Edges between components (defined by \verb|ECC|) are also rendered as solid blue lines.
  \item Dashed gray lines connect each original node to its corresponding component. The physics is disabled between nodes and their condensation components to help with readability.
\end{itemize}
 Figure~\ref{fig:graph-condensation} shows the result of using the above Logica code and the appropriate Python wrapper as described in Subsection~\ref{subsection:effortless} to view the original and corresponding condensed graph. 
By combining the power of Logica for defining graph structures and relationships with its seamless graph rendering capabilities, it is possible to gain valuable insights into complex systems through clear and concise visualizations.

\subsection{Inferring a Taxonomic Tree} 

Another application of Logica lies in its ability to infer relationships from large datasets and represent them as graphs. Consider the problem of constructing a simplified taxonomic tree for a set of species, showing their evolutionary relationships leading back to common ancestors. Logica can perform this task  on massive amounts of data.
Here, we aim to build a taxonomic tree for \textit{Homo sapiens} (humans), \textit{Crocodylidae} (crocodiles), \textit{Tyrannosaurus} (T-Rex), and \textit{Columbidae} (pigeons). The input is a knowledge graph of triples \texttt{T($a$,$b$,$c$)} and a functional predicate \texttt{L($a$)\,=\,$\ell$} that maps objects $a$ to human readable labels $\ell$. We define \texttt{SuperTaxon}(\emph{item},\emph{parent}) as the result of the query \texttt{T}(\emph{item},\texttt{"P171"},\emph{parent}), which states that \emph{parent} is a direct supertaxon of \emph{item}, and \texttt{TaxonLabel}($x$) = \texttt{L}($x$) to give labels restricted to taxons. The following Logica program builds the taxonomic tree.

\begin{mdframed}[backgroundcolor=gray!8!white,hidealllines=true, skipabove=0.2cm, innertopmargin=-0.0cm,innerbottommargin=-0.0cm] \begin{lstlisting}[style=logica]
# Use unbounded depth (-1) to find ancestors.
@Recursive(E, -1, stop: FoundCommonAncestor);
E(x, item,
  TaxonLabel(x), TaxonLabel(item)) distinct :-
  SuperTaxon(item,x),
  ItemOfInterest(item) | E(item);
NumRoots() += 1 :- E(x,y), ~E(z,x);
# Stop when common ancestor is found.
FoundCommonAncestor() :- NumRoots() = 1;
\end{lstlisting} 
\end{mdframed}

\begin{figure*}[t!]
  \centering
  \includegraphics[width=1.0\linewidth]{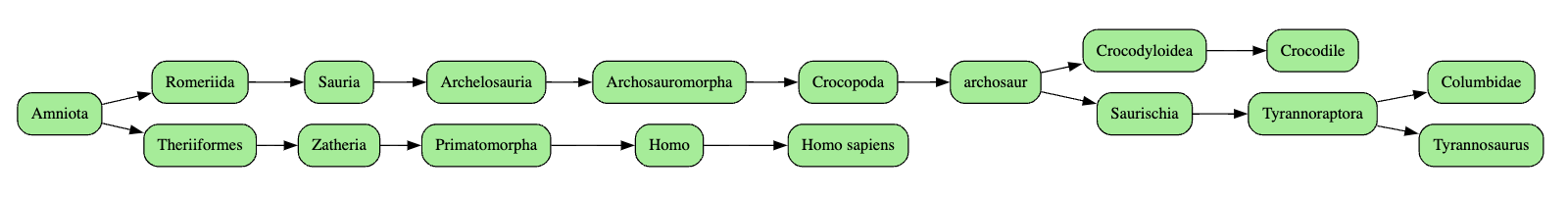}
  \caption{Inferred Taxonomic Tree for humans, crocodiles, T-Rex, and pigeons, generated by a Logica program from a large Wikidata dump. Edges represent evolutionary relationships, leading from an ancestral species to  descendant species. This demonstrates that Logica can operate on very large, real-world knowledge graphs.}
  \label{fig:taxonomic-tree}
\end{figure*}

Figure~\ref{fig:taxonomic-tree} shows an example output from running the program, where the GraphViz\footnote{\href{https://graphviz.org/}{https://graphviz.org/}}
library is used to render the resulting tree. The input for the example consisted of the set of statements obtained by a dump of Wikidata, which contained 806M facts defined over 89M objects. When stored using DuckDB, the entire input is 13GB in size.
The full recursive search was run on a Google Cloud \verb|c2d-standard-32| 
instance in under 7 seconds via DuckDB with no custom index setup. 
The majority of the execution time was spent selecting the taxonomy edges from all possible relations in Wikidata. Note that the number of taxons returned by the original program is large. The result shown in Figure~\ref{fig:taxonomic-tree} is only a sample of the obtained taxonomic tree (where the sampling is also performed by Logica; not shown above). 

By expressing the problem in Logica, we leverage its ability to efficiently perform (recursive) query processing (e.g., through its compilation to DuckDB). In this case, Logica is used to quickly navigate the complex relationships within a large taxonomic database and generate output that can then be displayed visually through its Python integration (here, using the GraphViz library).

\section{Conclusion}
We have presented a novel approach to implementing graph transformations using Logica, a free and open-source logic programming language designed for parallel databases. While Logica is typically used for data processing and analysis, we have shown that it can also effectively be used for graph transformations.
By leveraging Logica's ability to process large datasets in a parallel and efficient manner, graph transformations can be performed efficiently. By showcasing several examples, including the Win-Move problem, dynamic path finding, transitive reduction, graph condensation, and graph querying, we demonstrated that graph transformation problems can be expressed in an intuitive and efficient manner. The examples show how describing graph transformations using logic rules can lead to concise declarative specifications that also improve readability.  For example, our approach models time-varying graphs in a principled manner, an issue that can be challenging in classical graph transformation languages.

In future work, we plan to explore more complex graph transformation patterns, including rewritings that may require solving NP-hard problems. We also plan to benchmark our approach against other graph transformation tools, which would be valuable for demonstrating Logica's performance advantages on real-world datasets. 

We hope that Logica, with its ability to leverage underlying SQL engines, can make graph transformations a more widely accessible and practical approach for managing complex and dynamic data.


\bibliography{main}

\clearpage
\appendix


\end{document}

%% file: logica_style.tex
\usepackage{geometry}
\usepackage{textcomp}
\usepackage{listings}
\usepackage{xcolor}
\usepackage{filecontents}

\makeatletter

\newcommand\PrologPredicateStyle{}
\newcommand\PrologVarStyle{}
\newcommand\PrologAnonymVarStyle{}
\newcommand\PrologAtomStyle{}
\newcommand\PrologOtherStyle{}
\newcommand\PrologCommentStyle{}

\newif\ifpredicate@prolog@
\newif\ifwithinparens@prolog@

\lst@SaveOutputDef{`_}\underscore@prolog

\newcount\currentchar@prolog

\newcommand\@testChar@prolog%
{%
  \ifnum\lst@mode=\lst@Pmode%
    \detectTypeAndHighlight@prolog%
  \else
    \ifwithinparens@prolog@%
      \detectTypeAndHighlight@prolog%
    \fi
  \fi
  \global\predicate@prolog@false%
}

\newcommand\detectTypeAndHighlight@prolog
{%
  \def\lst@thestyle{\PrologAtomStyle}%
  \ifpredicate@prolog@%
    \def\lst@thestyle{\PrologPredicateStyle}%
  \else
    \expandafter\splitfirstchar@prolog\expandafter{\the\lst@token}%
    \expandafter\ifx\@testChar@prolog\underscore@prolog%
      \ifnum\lst@length=1%
        \let\lst@thestyle\PrologAnonymVarStyle%
      \else
        \let\lst@thestyle\PrologVarStyle%
      \fi
    \else
      \currentchar@prolog=65
      \loop
        \expandafter\ifnum\expandafter`\@testChar@prolog=\currentchar@prolog%
          \let\lst@thestyle\PrologVarStyle%
          \let\iterate\relax
        \fi
        \advance \currentchar@prolog by 1
        \unless\ifnum\currentchar@prolog>90
      \repeat
    \fi
  \fi
}
\newcommand\splitfirstchar@prolog{}
\def\splitfirstchar@prolog#1{\@splitfirstchar@prolog#1\relax}
\newcommand\@splitfirstchar@prolog{}
\def\@splitfirstchar@prolog#1#2\relax{\def\@testChar@prolog{#1}}

\def\beginlstdelim#1#2%
{%
  \def\endlstdelim{\PrologOtherStyle #2\egroup}%
  {\PrologOtherStyle #1}%
  \global\predicate@prolog@false%
  \withinparens@prolog@true%
  \bgroup\aftergroup\endlstdelim%
}

\newcommand\lang@prolog{Prolog-pretty}
\expandafter\lst@NormedDef\expandafter\normlang@prolog%
  \expandafter{\lang@prolog}

\expandafter\expandafter\expandafter\lstdefinelanguage\expandafter%
{\lang@prolog}
{%
  language            = Prolog,
  keywords            = {},      
  showstringspaces    = false,
  alsoletter          = (,
  alsoother           = @$,
  moredelim           = **[is][\beginlstdelim{(}{)}]{(}{)},
  MoreSelectCharTable =
    \lst@DefSaveDef{`(}\opparen@prolog{\global\predicate@prolog@true\opparen@prolog},
}

\newcommand\@ddedToOutput@prolog\relax
\lst@AddToHook{Output}{\@ddedToOutput@prolog}

\lst@AddToHook{PreInit}
{%
  \ifx\lst@language\normlang@prolog%
    \let\@ddedToOutput@prolog\@testChar@prolog%
  \fi
}

\lst@AddToHook{DeInit}{\renewcommand\@ddedToOutput@prolog{}}

\makeatother
\definecolor{PrologPredicate}{RGB}{000,031,255}
\definecolor{PrologVar}      {RGB}{024,021,125}
\definecolor{PrologAnonymVar}{RGB}{000,127,000}
\definecolor{PrologAtom}     {RGB}{186,032,032}
\definecolor{PrologComment}  {RGB}{063,128,127}
\definecolor{PrologOther}    {RGB}{000,000,000}
\definecolor{ColonKeywordColor}{RGB}{0,128,0}
\definecolor{EngineColor}{RGB}{139, 69, 19}

\renewcommand\PrologPredicateStyle{\color{PrologPredicate}}
\renewcommand\PrologVarStyle{\color{PrologVar}}
\renewcommand\PrologAnonymVarStyle{\color{PrologAnonymVar}}
\renewcommand\PrologAtomStyle{\color{PrologAtom}}
\renewcommand\PrologCommentStyle{\itshape\color{PrologComment}}
\renewcommand\PrologOtherStyle{\color{PrologOther}}

\lstdefinestyle{logica}
{
  language     = Prolog-pretty,
  upquote      = true,
  stringstyle  = \PrologAtomStyle,
  commentstyle = \PrologCommentStyle,
  basicstyle   = \footnotesize\ttfamily,
  comment      = [l]{\#},
  literate     =
    {:-}{{\PrologOtherStyle :- }}1
    {,}{{\PrologOtherStyle ,}}1
    {;}{{\PrologOtherStyle ;}}1
    {'}{{\PrologOtherStyle "}}1
    {.}{{\PrologOtherStyle .}}1
    {!}{{\PrologOtherStyle !}}1
    {~in~}{{\PrologOtherStyle in }}1
    {"}{{\PrologOtherStyle "}}1
    {[}{{\PrologOtherStyle [}}1
    {]}{{\PrologOtherStyle ]}}2
    {true}{{\PrologOtherStyle true}}1
    {false}{{\PrologOtherStyle false}}1
    {nil}{{\PrologOtherStyle nil}}1
    {"to"}{{\PrologOtherStyle "to" }}1
    {"arrows"}{{\PrologOtherStyle "arrows" }}1
    {"physics"}{{\PrologOtherStyle "physics" }}1
    {"dashes"}{{\PrologOtherStyle "dashes" }}1
    {"smooth"}{{\PrologOtherStyle "smooth" }}1
    {"color"}{{\PrologOtherStyle "color" }}1
    {"width"}{{\PrologOtherStyle "width" }}1
    {"\#33e"}{{\PrologOtherStyle "\#33e"}}1
    {"\#888"}{{\PrologOtherStyle "\#888"}}1
    {"c-"}{{\PrologOtherStyle "c-"}}1
    {rgba}{{\PrologOtherStyle rgba }}1
    {fact.order\_id}{{\PrologAtomStyle fact.order\_id}}1
    {graph.SimpleGraph}{{\PrologAtomStyle graph.SimpleGraph}}1
    {Report}{{\PrologPredicateStyle Report}}1
    {distinct}{{{\color{EngineColor} distinct}}}1
    {@Recursive}{{{\color{EngineColor} @Recursive}}}1
    {@Engine}{{{\color{EngineColor} @Engine}}}1
    {creator:}{{{\color{ColonKeywordColor} creator:}}}7
    {t0:}{{{\color{ColonKeywordColor} t0:}}}7
    {t1:}{{{\color{ColonKeywordColor} t1:}}}7
    {color:}{{{\color{ColonKeywordColor} color:}}}7
    {arrows:}{{{\color{ColonKeywordColor} arrows:}}}7
    {color?}{{{\color{ColonKeywordColor} color?}}}7
    {dashes?}{{{\color{ColonKeywordColor} dashes?}}}7
    {physics?}{{{\color{ColonKeywordColor} physics?}}}7
    {smooth?}{{{\color{ColonKeywordColor} smooth?}}}7
    {width?}{{{\color{ColonKeywordColor} width?}}}7
    {physics:}{{{\color{ColonKeywordColor} physics:}}}7
    {dashes:}{{{\color{ColonKeywordColor} dashes:}}}7
    {smooth:}{{{\color{ColonKeywordColor} smooth:}}}7
}